\documentclass[aip,prl,preprint,onecolumn,groupedaddress]{revtex4-1}
\usepackage{amsmath}	%need for subequations
\usepackage{setspace}	%need for double spacing
\usepackage{graphicx}	%need for figures
\usepackage{subfigure}	%use for side-by-side figures
\usepackage{color}		%use if color is used in text
\usepackage{verbatim}	%use comment command
\graphicspath{{./figs/}}
\usepackage{chemformula} % Formula subscripts using \ch{}
\usepackage[T1]{fontenc} % Use modern font encodings
\usepackage{subfigure}
\usepackage{bm}

\begin{document}

%\preprint{}
\title{Rebound suppression of a droplet impacting on a supersolvophobic surface by a small amount of polymer additives}


\author{Eunsang Lee}
\email[Electronic mail:]{e.lee@theo.chemie.tu-darmstadt.de}
\affiliation{Eduard-Zintl-Institut f\"ur Anorganische und Physikalische Chemie, Technische Universit\"at Darmstadt, Alarich-Weiss-Stra{\ss}e 8, 64287 Darmstadt, Germany}
\author{Hari Krishna Chilukoti}
\thanks{Current affiliation: Department of Mechanical Engineering, National Institute of Technology Warangal, Warangal, Telangana 506004, India}
\affiliation{Eduard-Zintl-Institut f\"ur Anorganische und Physikalische Chemie, Technische Universit\"at Darmstadt, Alarich-Weiss-Stra{\ss}e 8, 64287 Darmstadt, Germany}
\author{Florian M\"uller-Plathe}
\affiliation{Eduard-Zintl-Institut f\"ur Anorganische und Physikalische Chemie, Technische Universit\"at Darmstadt, Alarich-Weiss-Stra{\ss}e 8, 64287 Darmstadt, Germany}

%used with the \author command)

\date{\today}
\begin{abstract}

A small amount of polymer dissolved in a droplet suppresses droplet rebound when it impinges on a supersolvophobic surface.
This work investigates impacting dynamics of a droplet of dilute polymer solution depending on the molecular weight and the concentration of the polymer by using multi-body dissipative particle dynamics simulations.
Either the longer polymer or the high polymer concentration suppresses rebound of a droplet although its shear viscosity and the liquid-vapor surface tension are not different from those of a pure solvent droplet.
We found a new mechanism of the anti-rebound in which the resistance is applied against the hopping motion, while behavior of the non-rebounding droplet at the earlier spreading and retraction stages is same as for the rebounding droplets.
Two polymer contributions to reducing the rebound tendency are quantitatively analyzed: the alteration of the substrate wettability by the polymer adsorption and the polymer elongation force. 

\end{abstract}

%\pacs{}
%\keywords{}

\maketitle


A droplet of a Newtonian liquid which impinges on a highly solvophobic surface, will normally rebound, at least partially. 
This rebound is reliably suppressed by the addition of a small amount of polymer, for example, 200 ppm of polyethylene oxide (PEO) in water.\cite{bertola2013} 
The anti-rebound behavior of a dilute polymer solution droplet is of great scientific and practical interest, but its underlying physics is poorly understood.\cite{wirth1991, monteux2008} 
The fundamental interest arises particularly from the observation that adding very low concentration of polymer to a droplet does not alter its shear viscosity and liquid-vapor surface tension, but leads to a dramatic change in its tendency to rebound.
This contradicts the traditional explanation of spreading and retraction processes of an impacting droplet in terms of a balance between inertia, viscous force, and capillary force. 
They are typically characterized by the Reynolds number, Re = $D_{0}v_\mathrm{imp}\rho/\eta_\mathrm{s}$ and the Weber number, We = $D_{0}v_\mathrm{imp}^2\rho/\gamma$, where $D_0$ is the equilibrium droplet diameter, $v_\mathrm{imp}$ is the impact velocity, $\rho$ is the droplet mass density, $\eta_\mathrm{s}$ is the shear viscosity, and $\gamma$ is the fluid's liquid-vapor surface tension.\cite{pasandidehfard1996,josserand2016,ukiwe2005,attane2007,lee2016}
However, there is no concrete relation between these numbers and the rebound tendency of a dilute polymer solution droplet.
This calls for new theoretical and numerical approaches.

Aided by the development of high speed imaging techniques, an early attempt to understand the polymer effect on the rebound tendency was carried out.
It suggested that non-Newtonian elongational viscosity brought by the polymer leads to a reduced retraction velocity of a droplet.\cite{bergeron2000}
This argument, however, has been debated, because the liquid is under shear flow, not under elongational flow, during the retraction stage.
A later study attributed the rebound suppression to the increased friction of the three-phase contact line during the retraction stage, imposed by polymer deposition on the substrate.\cite{zang2013}
This view was also supported by stick-slip dynamics of the contact line and lamella-structured polymer deposit left on the substrate.\cite{smith2010, bertola2013}
An experiment demonstrating that asymmetrical pre-deposition of polymer on a substrate slows the retraction speed, with the consequence of an asymmetric rebound, also highlighted the role of the additional friction provided by the deposited polymer.\cite{huh2015}
Most recently, the importance of the elongational viscosity was revisited by a high damping coefficient of height oscillation of a droplet of PEO in water after impact.\cite{chen2018}
However, the interpretation of the rebound in above studies is limited to the phenomenology and it lacks a molecular picture.
No direct evidence of a polymer effect on the anti-rebound has been provided by numerical solutions of the Navier-Stokes equation with finite-element nonlinear elastic polymer models.\cite{izbassarov2016,tembely2019} 
This has been attributed to the inability to describe the wettability change of a substrate and the variation of local polymer concentration within the droplet.
A more suitable way to find the physics behind the anti-rebound effect is a particle-based simulation, but to our best knowledge, no such studies have been reported.
In this letter, we investigate the anti-rebound behavior of an impacting droplet of a dilute polymer solution, while varying systematically the molecular weight and the concentration of the polymer by using multi-body dissipative particle dynamics (MDPD) simulations. 

The central idea of the original dissipative particle dynamics (DPD) simulation is to reproduce correct hydrodynamics by viscous forces between groups of atoms.\cite{espanol1995,espanol2017}
In DPD, a large integration time step is possible due to the soft nature of purely repulsive potential functions, thus mesoscopic length and time scales are achievable. 
Recently, keeping all the benefits of the DPD method, the limitation of the invariable quadratic equation of state of DPD was overcome by the addition of a local density-dependent multi-body attractive force, which is named MDPD.\cite{warren2003}
The MDPD allows liquid-vapor coexistence and generates a sharp interface, which is necessary for the droplet simulations in this work.
We investigate droplets of six different polymer molecular weights (number of polymer beads $N_\mathrm{p}$= 1--100), and of five different number fractions of polymer beads ($x_\mathrm{p}$= 0.5--2.5 $\times 10^{-2}$) for each $N_\mathrm{p}$.
The highest polymer concentration is only one tenth of the overlap concentration even for $N_\mathrm{p}$= 100.
Thus, our solutions are in the very dilute regime.
Each droplet contains 4$\times 10^4$ beads including polymer and solvent beads.
The substrate (surface) is modeled by frozen particles on a square lattice.
The repulsion amplitude and the cutoff distance are chosen to be same for all particle pairs ($B=25$, $r_B=1.0$). (See Supplemental Material for details of simulation methods and models, including DPD reduced units) 
The amplitudes of the attractive force ($A$) differ depending on the pair type such that $A_\mathrm{s/s}=A_\mathrm{p/p}=A_\mathrm{s/p}=-40$, $A_\mathrm{s/w}=-10$, and $A_\mathrm{p/w}=-30$, where subscripts s, p, w refer to solvent, polymer and substrate (``wall'') beads, respectively. 
Such a choice of interaction parameters at temperature $k_\mathrm{B}T=1$ mimics polymer in a good solvent, and a solvophobic but polymerphilic substrate. 
The equilibrium contact angle for a pure solvent droplet on the substrate is 155\textdegree.
After equilibrating a free droplet, we let it impact on the substrate with a constant velocity of 2.
The surface tension of free droplets and the zero-shear viscosity of corresponding polymer solutions increase slightly as either $N_\mathrm{p}$ or $x_\mathrm{p}$ increases, but without meaningful difference beyond the statistical error.
Together with equilibrium droplet diameters ($D_{0}=23.2$) and liquid mass densities ($\rho=6.10$) of droplets, which are nearly constant irrespective of polymer composition, the Reynolds and Weber numbers are calculated to be similar as Re= 35--42 and We= 78--81. (See Supplemental Material)

\begin{figure*}[t]
\centering
\includegraphics[width=1.0\textwidth]{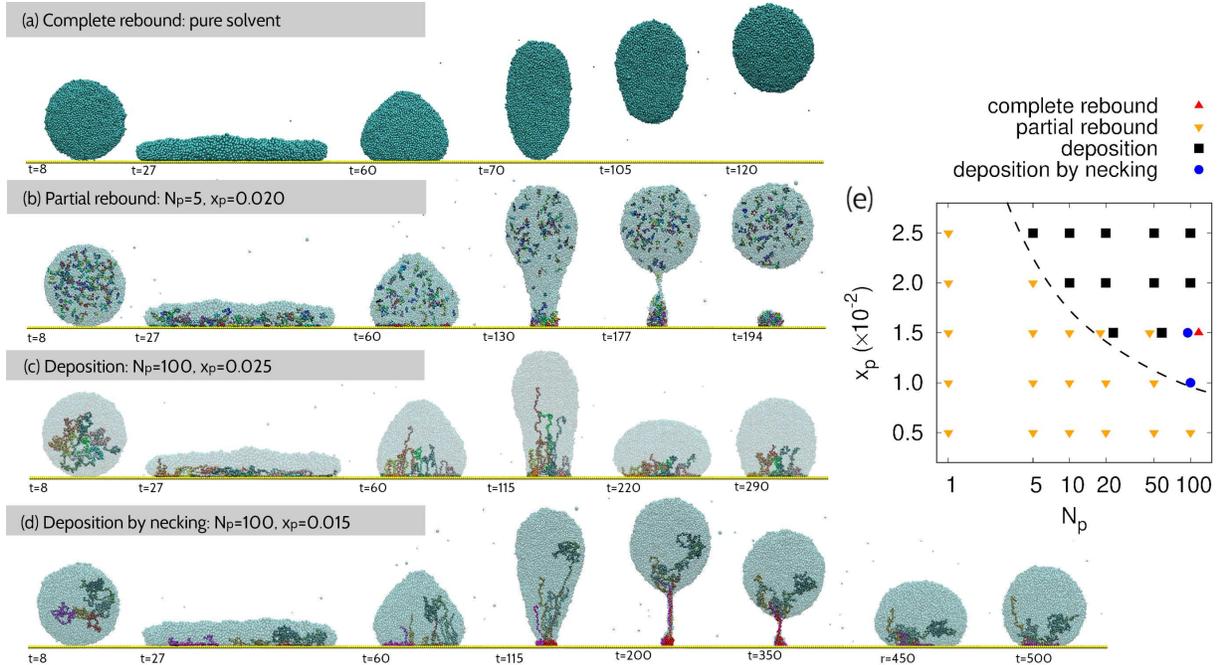}
\caption{\label{fig-outcomes}Simulation snapshots of four different outcomes from side view: (a) complete rebound, (b) partial rebound, (c) deposition, and (d) deposition by polymer necking. In all images, solvent and substrate particles are colored in cyan and yellow, respectively. Each polymer strand is depicted in different color. In Figure (e), the most probable outcomes are shown for different polymer molecular weight ($N_\mathrm{p}$) and concentration ($x_\mathrm{p}$). Two outcomes for a given $N_\mathrm{p}$ and $x_\mathrm{p}$ (for example, deposition and partial rebound for $N_\mathrm{p}$= 20, $x_\mathrm{p}$= 0.015) indicate that two outcomes are equally probable out of five independent trajectories. The black dashed line on the graph is a rough boundary between rebound and deposition. Snapshots are obtained by VMD.\cite{humphrey1996}}
\end{figure*}

Right after impact the droplet first spreads. 
After reaching a maximum expansion, it retracts back toward the droplet center followed by four different outcomes of rebound: complete rebound, partial rebound, deposition, and deposition by polymer necking as displayed in Figures \ref{fig-outcomes}(a)--(d). (See also movies in Supplemental Material)
Figure \ref{fig-outcomes}(e) shows the most probable outcome out of five independent trajectories for each polymer composition. 
A droplet of pure solvent rebounds completely resulting from high enough inertia and surface tension as intended by chosen simulation parameters. 
Droplets of small $x_\mathrm{p}$ and small $N_\mathrm{p}$ (left lower part of Figure \ref{fig-outcomes}(e)) partially rebound with a small satellite droplet left on the substrate, which was also observed in several experiments.\cite{bertola2013,huh2015}
The satellite drop is enriched in polymer of the number fraction larger than 0.5.
As in experiments, we find a general trend that droplets containing polymer of longer $N_\mathrm{p}$ and higher $x_\mathrm{p}$ do not rebound, although their solution viscosity is not much larger than that of the pure solvent. 
A key factor of rebound suppression, which can be extracted from this figure is that polymer molecules are largely adsorbed on the substrate to mediate the unfavorable interaction between substrate and solvent. 
Moreover, the remaining part of the polymer not adsorbed on the substrate is stretched normal to the substrate, i.e. the direction of rebound, with the flowing solvent.
After the polymer is highly stretched, it tends to recover its coiled structure, exerting a force against the rebound to the surrounding solvent particles.
The effect of the stretched polymer is most prominent in the final outcome, namely deposition by polymer necking (Figure 1(d)).
This is observed especially for the longest polymer of $N_\mathrm{p}=$ 50 or 100 at an intermediate concentration of $x_\mathrm{p}=$ 0.010 or 0.015.
As only some polymer beads are adsorbed on the substrate, the majority of a long polymer strand can stretch in the direction of rebound. 
Most solvent particles temporarily leave the substrate and form a partial droplet above the substrate.
The polymer, however, whose one end is adsorbed on the substrate and the other end is buried in the solvent droplet, actually pulls the droplet back while recovering its coiled structure ($t=$ 200). 
As a result, the whole droplet ends up deposited on the substrate.
Such a polymer necking conformation has not been found or postulated in any droplet impact experiments, but a droplet shape very similar to our simulation snapshot at $t=$ 200 in Figure \ref{fig-outcomes}(d) has been observed in experiments using a high-speed camera.\cite{huh2015}
A stretched polymer configuration and the necking morphology observed in droplet pinch-off experiments\cite{sachdev2016} also strongly support the reality of the formation of a polymer necking morphology during droplet impact.

\begin{figure}[t!]
\centering
\subfigure[]{\label{fig-spreading}\includegraphics[width=0.48\textwidth]{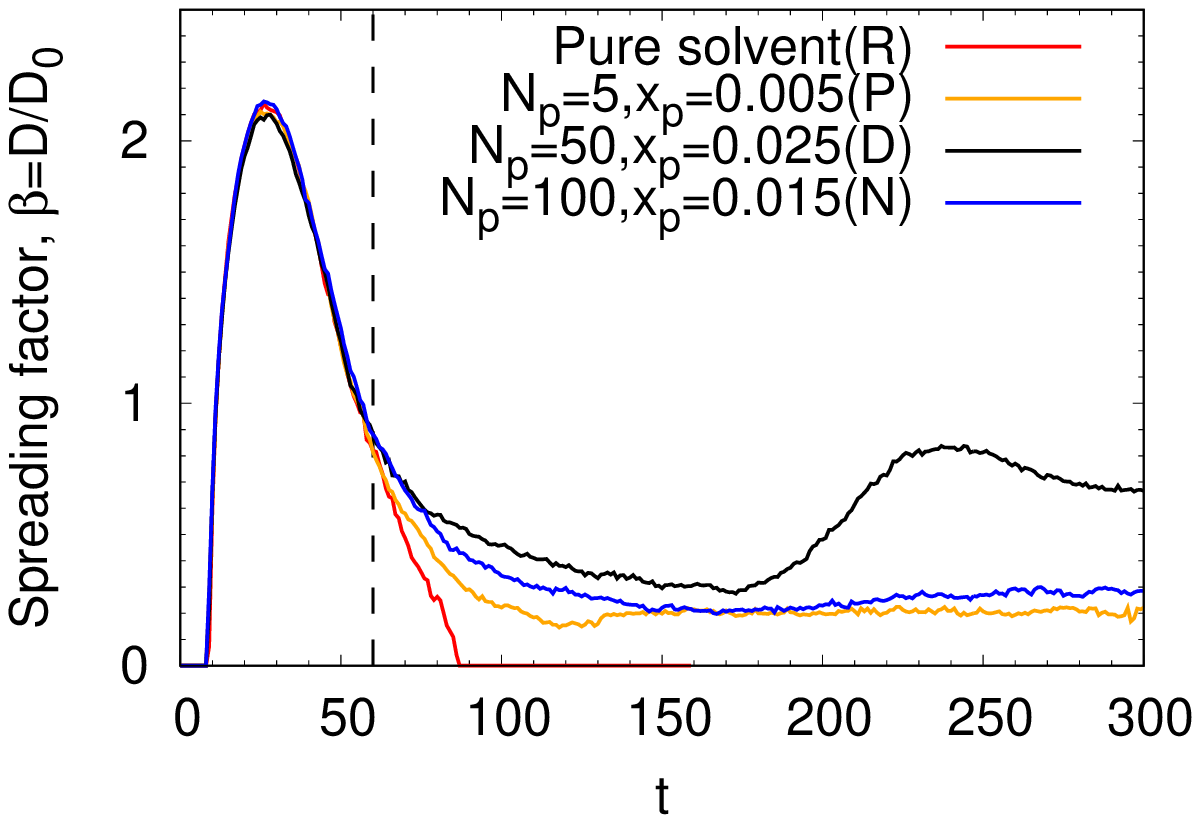}}
\subfigure[]{\label{fig-velocity}\includegraphics[width=0.48\textwidth]{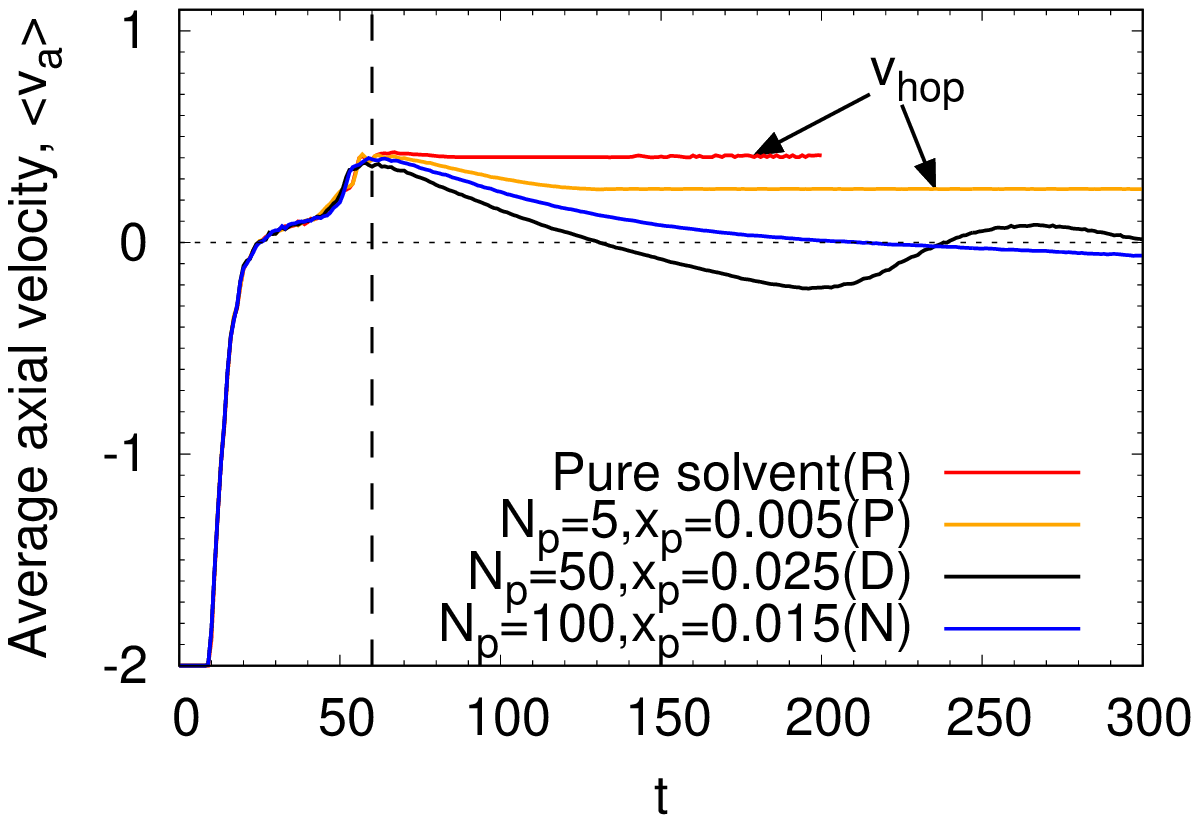}}

\caption{\label{fig-sprvel} (a) Spreading factor and  (b) average particle axial velocity as a function of time for droplets of different outcomes: complete rebound (R), partial rebound (P), deposition (D) and deposition by necking (N). Vertical dashed lines indicate the beginning of a hopping stage at $t=60$. Arrows in Figure (b) indicate the hopping velocity.}
\end{figure}

The spreading diameter $D$ is the diameter of the circle formed by the three-phase contact line.
It is shown as the spreading factor, $\beta=D/D_0$ as a function of time in Figure \ref{fig-spreading}. 
The maximum spreading factors, $\beta_\mathrm{max}$, of all droplets are almost the same independent of polymer composition because they have almost the same We and Re numbers.\citep{ukiwe2005} 
This is also evident in the snapshots at $t=$ 27 of Figure \ref{fig-outcomes}(a)--(d).
Figure \ref{fig-spreading} shows that the velocity of the three-phase contact line is reduced for non-rebounding droplets after $\beta_\mathrm{max}$ is attained.
However, there is a striking difference between our simulation and experimental results: In the simulation, the slowing of the contact line motion begins at the very end of the retraction stage, $t\approx 60$ (vertical dashed lines in Figure \ref{fig-sprvel}) when $\beta\approx 1$, while in experiments the retraction slows down right after $\beta$ reaches $\beta_\mathrm{max}$.\cite{bergeron2000, bertola2013, huh2015}
This result is consistent with the fact that droplets, which later will have different rebounding outcomes, still show almost the same morphology and size at early times up to $t=$ 60 in Figures \ref{fig-outcomes}(a)--(d).
This behavior is more clearly seen in the average particle velocity in the direction normal to the surface in Figure \ref{fig-velocity}.
This figure shows that particles in the non-rebounding droplet also reach an axial velocity as high as in the rebounding droplets at $t\approx 60$.
Only after that time, the velocity decreases depending on the strength of resistance against the rebound motion.
Looking at the shape of the droplets at $t=$ 60 in Figure \ref{fig-outcomes}(a)--(d), they are no longer compressible in the surface plane and the particles' momentum starts to be diverted from parallel to perpendicular direction to the substrate. (See also Supplemental Material for almost zero radial particle velocities at $t>60$)
Therefore, the rebound of the droplets is suppressed not by the resistance against the parallel retraction but by the resistance against the perpendicular motion.
We newly name this stage of $t>60$ as a \textit{hopping stage} and will focus on this stage in the rest of this letter.
One should keep in mind that our finding does not deny the experimental observations but rather states that the rebound can also be suppressed during the hopping stage, even though the retraction velocity is the same as for the rebounding droplet.

\begin{figure}[t!]
\subfigure[]{\label{fig-normalstress}\includegraphics[width=0.48\textwidth]{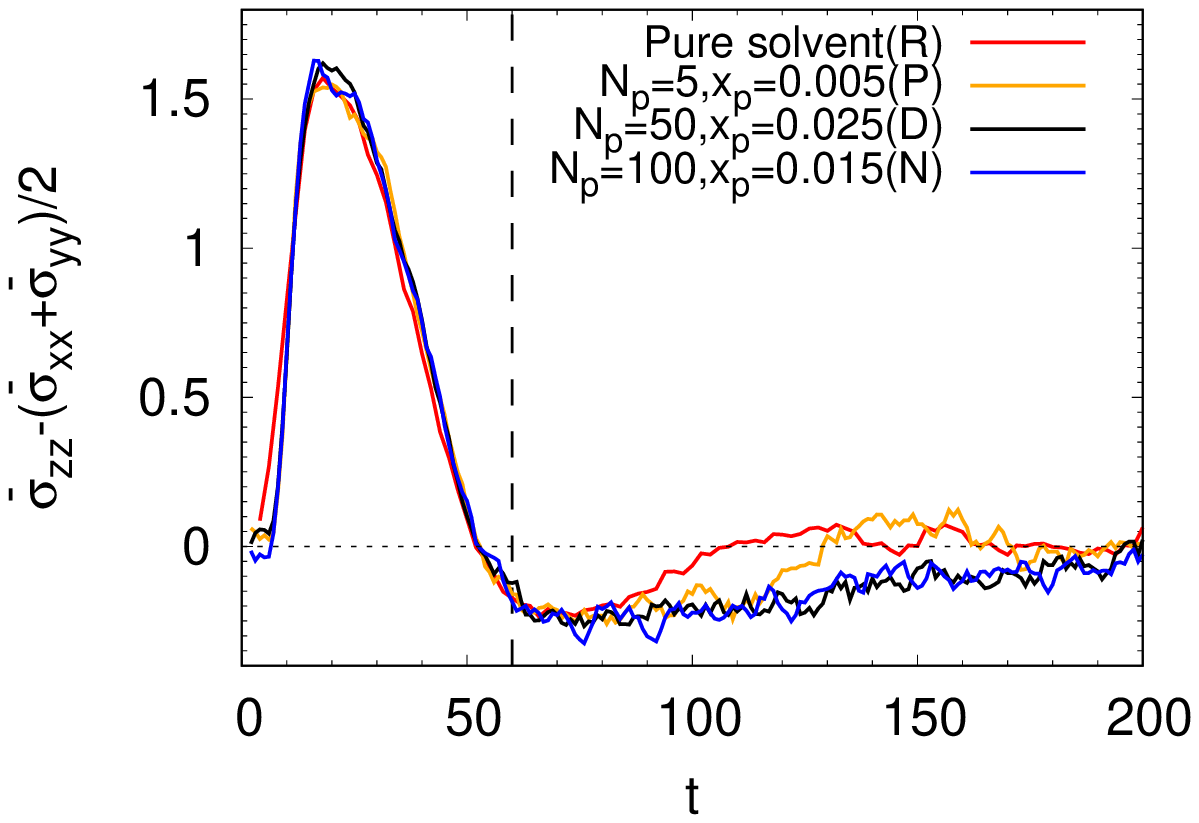}}
\subfigure[]{\label{fig-contangle}\includegraphics[width=0.48\textwidth]{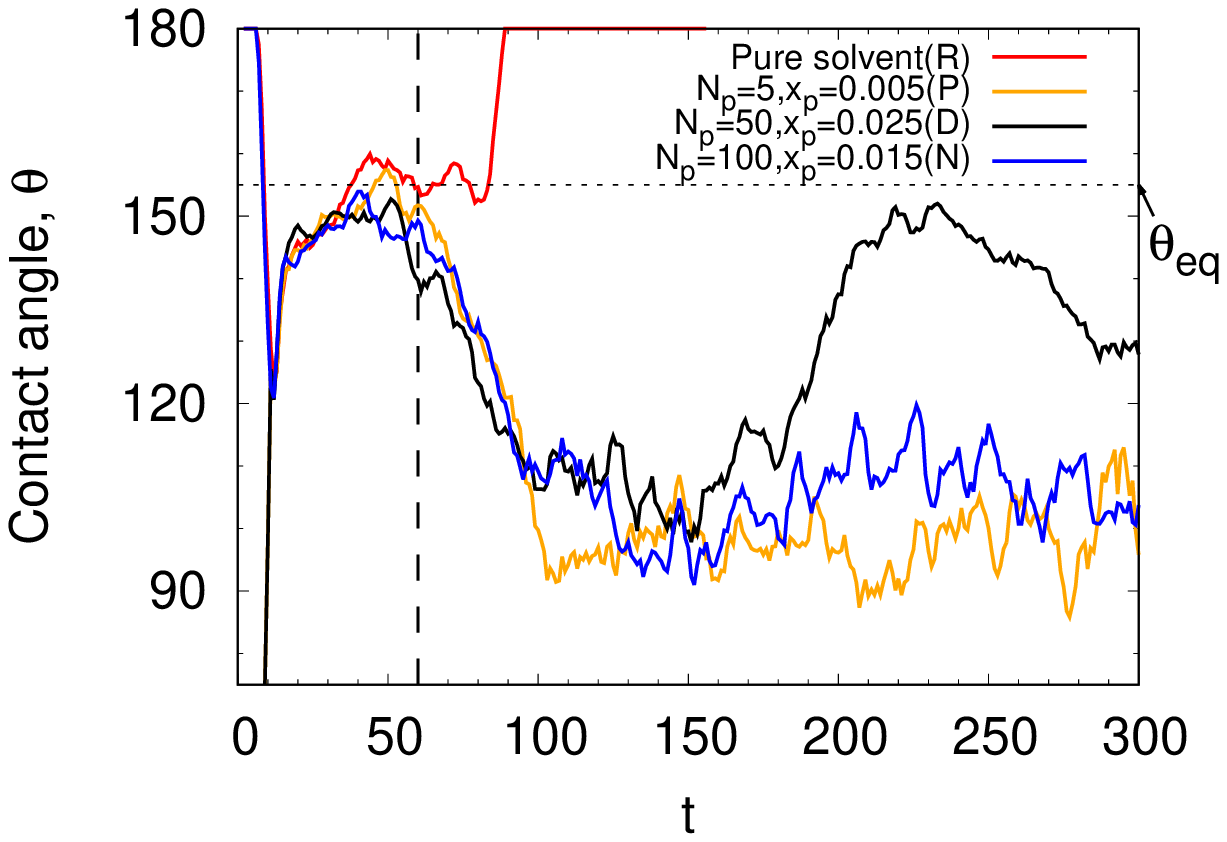}}
\subfigure[]{\label{fig-polads}\includegraphics[width=0.48\textwidth]{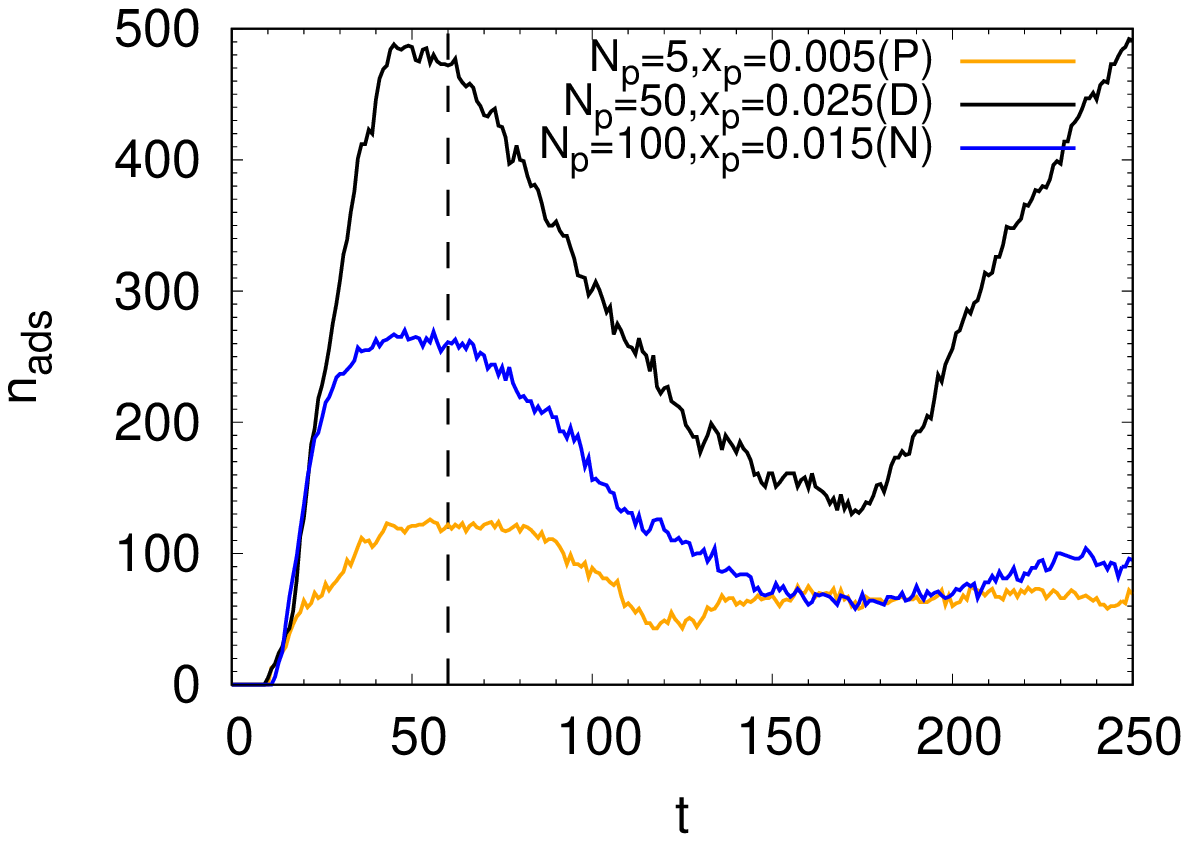}}
\caption{\label{fig-polcont} (a) First normal stress difference, (b) contact angle, and (c) the number of adsorbed polymer beads as a function of time for droplets of different outcomes. Vertical dashed lines indicate the beginning of a hopping regime at $t=60$. 5-point running averages of the contact angle for better statistics are plotted in (b). Horizontal black dotted lines in (a) and (b) indicate 0 and the equilibrium contact angle of the pure-solvent droplet, $\theta_\mathrm{eq}=155$, respectively.}
\end{figure}

To monitor the force that droplet beads feel along the perpendicular direction to the substrate ($z$-direction), we calculate the first normal stress difference, $N_1=\sigma_{zz}-(\sigma_{xx}+\sigma_{yy})/2$ in Figure \ref{fig-normalstress}.
This Figure shows $N_1$ obtained from 5-point running averages of stress tensor components.
Consistent with the finding in Figure \ref{fig-sprvel}, the main difference in $N_1$ between different rebounding outcomes appears during the hopping stage.
The negative stress in this stage represents a net attractive force between particles and its amount is bigger for non-rebounding droplets than for rebounding droplets especially at large $t>60$. 
As the elongational viscosity, $\eta_\mathrm{e}$, of a liquid under uniaxial elongational flow is given by $\eta_\mathrm{e}=-N_1 / \dot{\epsilon}$, where $\dot{\epsilon}$ is an elongational flow rate, the negative $N_1$ is mainly contributed by the elongational force of polymer.
However, even the pure-solvent droplet has a negative stress at the early hopping stage because the net attractive force primarily stems from interaction at the liquid-solid interface.(Solvent and substrate beads also weakly attract each other.)
At the interface, if polymer beads are adsorbed on the substrate as shown in Figure \ref{fig-outcomes}, they attract both solvent and substrate beads to mediate unfavorable solvent-substrate interaction, leading to large $|N_1|$.

A time evolution of the contact angle shows the effect of deposited polymer on the dynamic wetting. (See Supplemental Material for the method)
We find $\approx$10\textdegree~ difference between rebounding and non-rebounding droplets in the retracting (receding) contact angle at $t=60$, which is similar to the experimental observation.\cite{bertola2015} 
However, a big difference is again shown in the hopping stage.
While the contact angle of the completely rebounding droplet stays $\approx$ 155\textdegree~ until it detaches, the contact angle of non-rebounding droplets decreases to $\approx$ 90\textdegree~ until $t\approx 100 $ which is an equilibrium contact angle of a droplet containing only polymer beads. 
The decreased contact angle during the hopping stage means strong attractive forces at the liquid-solid interface.
Taking an advantage of particle-based simulations, we calculate the number of polymer beads adsorbed on the substrate ($n_\mathrm{ads}$) as a function of time (Figure \ref{fig-polads}), which quantifies the change of the wettability of the substrate.
We define as adsorbed polymer beads those, whose distance from the substrate is less than 1, the cutoff distance of the attractive force. 
Previous numerical studies showed that there is no universal behavior of polymer adsorption and desorption under shear flow, and both depend strongly on the flow rate and the polymer-substrate interaction.\cite{dutta2013, dutta2015}
Under our specific condition $n_\mathrm{ads}$ increases during the spreading and early retraction stages, and stays constant at the late retraction stage ($t=50-60$) due to the balance between flow rate and the substrate-polymer attraction. 
At the beginning of the hopping stage ($t=60)$) the liquid flow direction changes to the substrate normal, and $n_\mathrm{ads}$ starts to decrease. 
Therefore, we take the maximum value of $n_\mathrm{ads}$, $n_\mathrm{ads,max}$, to characterize the change of the substrate wettability by the polymer adsorption.
As expected, we obtain larger $n_\mathrm{ads,max}$ for longer $N_\mathrm{p}$ and larger $x_\mathrm{p}$, (See Supplemental Material) which represents the stronger attractive force acting on a droplet during the hopping stage.

\begin{figure}[t]
\centering
\includegraphics[width=0.48\textwidth]{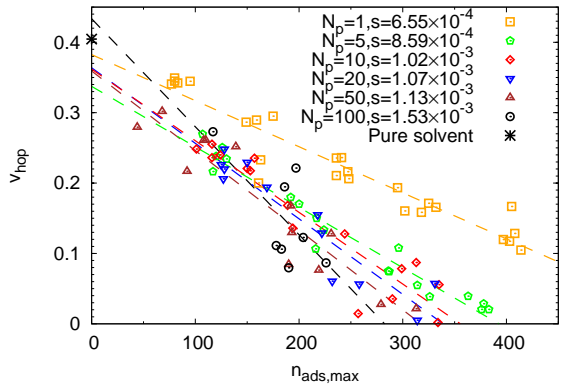}
\caption{\label{fig-hopvel} Hopping velocity versus the maximum value of the number of adsorbed polymer beads for different $N_\mathrm{p}$ and $x_\mathrm{p}$. Each $v_\mathrm{hop}$ is obtained from each trajectory in which the droplet rebounds. Fitted linear functions for each $N_\mathrm{p}$ ($v_\mathrm{hop}=-sn_\mathrm{ads,max}+c$) are represented by dashed lines, where the fitted $s$ are given in the legend. }
\end{figure}

To quantitatively understand the effect of polymer adsorption on the rebound behavior, we plot the hopping velocity, $v_\mathrm{hop}$, as a function of $n_\mathrm{ads,max}$ in Figure \ref{fig-hopvel}.
Here, $v_\mathrm{hop}$ is defined as the constant velocity after the droplet detaches from the substrate (Figure \ref{fig-velocity}).
As expected, $v_\mathrm{hop}$ decreases as more polymer beads are adsorbed on the substrate for a given $N_\mathrm{p}$. 
We can fit the data for each $N_\mathrm{p}$ with a line (dashed lines), especially for short polymers. 
%For longer polymers, trajectories fluctuate more giving a wider distribution of hoping velocities, but they are still scalable by a linear function. 
It is noticeable that the slope monotonously decreases with increasing $N_\mathrm{p}$. 
In other words, for a given $n_\mathrm{ads,max}$ characterizing a given substrate wettability, a droplet containing longer polymer chains is less likely to rebound. 
This finding clearly indicates that the polymer elongation force additionally affects the rebound behavior, as already expected from Figures \ref{fig-outcomes} and \ref{fig-normalstress}.
The elongation force originates from the tendency of the polymer to recover its coiled equilibrium conformation, and it generates a pulling force on the droplet by hydrodynamic interaction.
Assuming Hookean elasticity of the stretched polymer in infinite dilution, the recovering force is proportional to the end-to-end distance or alternatively the stretching factor of the stretched polymer.
We find that the stretching factor $\langle S_z \rangle$, which is the difference along the $z$-direction between uppermost and lowermost beads of a chain, is roughly proportional to $N_\mathrm{p}$ because the polymer is almost fully stretched. (See Supplemental Material)
This indicates that the elongation force per chain also scales as $N_\mathrm{p}$.
However, the number of polymers for a given $x_\mathrm{p}$ is inversely proportional to $N_\mathrm{p}$, leading to the same total elongation force under assumptions of no interchain interactions and a homogeneous polymer concentration.
The main difference in the resistance comes from how long the elongation force acts on the droplet which is again related to the amount of stretching.
This can be also observed in Figure \ref{fig-normalstress} where the minimum value of the negative stress during the hopping stage is almost the same for all non-rebounding droplets and the main difference is how long the negative stress lasts. 
As a result, a higher polymer molecular weight leads to a stronger total resistance against rebound.
The slope of the fitted functions ($s$) in Figure \ref{fig-hopvel} scales as $\sim -N_\mathrm{p}^{0.17(3)}$. (See Supplemental Material for a graph of $v_\mathrm{hop}$ as a function of $n_\mathrm{ads,max}\cdot N_\mathrm{p}^{0.17}$)
However, a droplet containing longer polymers has a larger polymer concentration locally near the neck than of shorter polymer,  which makes it even more complicated to find a relation. 
We are now investigating the rebound behavior depending on the elongational viscosity of the polymer solutions by considering the local polymer concentration and the stretching of the unadsorbed polymer strand during the hopping stage.

To conclude, we studied the rebound of droplets of polymer solution from a solvophobic substrate by MDPD simulations. 
We found that the droplet rebound is suppressed by polymer after the end of the retraction stage when the main flux has been converted from parallel to perpendicular to the substrate. 
While the velocity of the three-phase contact line during the retraction stage is not reduced, the polymer manifests itself during this hopping stage. 
It makes the two polymer contributions to reducing the rebound tendency: The first is to alter the liquid-solid interfacial interaction by adsorbed polymer primarily generating an attraction.
The second is the restoring force of the polymer after it has been stretched in perpendicular direction.
The alteration of the liquid-solid interface is characterized by the number of polymer beads adsorbed on the substrate that shows an anti-correlation with the hopping velocity. 
For a given substrate wettability modified by polymer adsorption, a long polymer pulls the droplet back to the substrate for a long time because it is highly stretched. 
Thus, we find that the polymer counteracts droplet rebound at the stage, when the droplet wants to detach from the substrate, rather than early, when the droplet retracts from its maximum lateral expansion.
While our interpretation is consistent with the fact that a small polymer content does not significantly changes the viscosity and the surface tension of the liquid, we cannot rule out that other mechanisms might prevail for different combinations of material's properties, such as solvophilicities vs. adsorptivities. 
At the same time, our results open the way to tailoring polymer additives against droplet rebound.

\begin{acknowledgements}

We would like to thank for the financial support by the Collaborative Research Centre Transregio 75 (project number 84292822) 
\textit{Droplet Dynamics in Extreme Environments} of the Deutsche Forschungsgemeinschaft. Calculations for this research were conducted on the Lichtenberg high performance computer of the TU Darmstadt and the GCS Supercomputer JUWELS at J\"ulich Supercomputing Centre (JSC) through the John von Neumann Institute for Computing (NIC).

\end{acknowledgements}

%\appendix
%\section{}

%\bibliographystyle{jabbrv}
\bibliography{droplet_prl}

\end{document}